\documentclass[prb,twocolumn,groupedaddress,shownopacs,amssymb]{revtex4-2}
\usepackage{graphicx,psfrag,amsmath,amssymb}

\usepackage[usenames, dvipsnames]{color}

\usepackage{hyperref} 
\usepackage[normalem]{ulem}
\usepackage{xcolor}
\hypersetup{
    colorlinks=true,
    linkcolor=Blue,
    citecolor=Blue,
    filecolor=Blue,      
    urlcolor=Blue,
    pdftitle={allflatband paper},
    }
\begin{document}

\title{Correlation lengths of flat-band superconductivity from quantum geometry}
\author{S. S. Elden and M. Iskin}
\affiliation{
Department of Physics, Ko\c{c} University, Rumelifeneri Yolu, 
34450 Sar\i yer, Istanbul, T\"urkiye
}

\date{\today}

\begin{abstract}

Flat-band superconductors provide a regime in which kinetic energy is
quenched, so that pairing is governed primarily by interactions and
quantum geometry. We investigate characteristic superconducting length
scales in all-flat-band systems under the assumptions of time-reversal
symmetry and spatially-uniform pairing, focusing on the size of the
lowest-lying two-body bound state, the average Cooper-pair size, and the
zero-temperature coherence length in two-band Hubbard models. Using the
Creutz ladder and the $\chi$ lattice as representative examples, we show
that both the two-body bound-state size and the many-body Cooper-pair
size remain finite and small in the weak-coupling limit, being controlled
by the quantum metric of the flat bands. By contrast, the coherence
length exhibits qualitatively distinct behavior, diverging in the dilute
limit and in the vicinity of insulating regimes. These results
demonstrate that, in flat-band superconductors, the pair size and the
coherence length are fundamentally distinct physical quantities and
highlight the central role of band geometry in shaping
superconducting length scales when kinetic energy is quenched.

\end{abstract}
\maketitle

\section{Introduction}
\label{sec:intro}

Flat-band superconductivity represents a profound departure from 
conventional BCS theory, as pairing can emerge even in the complete 
absence of kinetic-energy dispersion, rendering interaction 
effects and the geometry of Bloch states central to the 
superconducting phenomenology~\cite{torma22, peotta23, yu25, liu25, gao25}. 
Early studies established that flat bands can support superconductivity
with a critical temperature that scales linearly with the interaction
strength~\cite{kopnin11}; subsequent work has shown, however, 
that superconducting coherence is governed not by the density of 
states alone but by the quantum geometry of the underlying 
band structure~\cite{torma22, peotta23, yu25}. 
In particular, the quantum metric tensor, which quantifies distances
between Bloch states in momentum space, becomes especially relevant
when band dispersion is quenched and conventional
dispersion-based mechanisms for superconductivity are absent.
Until very recently, the quantum metric could not be accessed
experimentally in solid-state systems; however, two independent
breakthroughs have now shown that quantum geometry is directly
observable in crystalline materials, rather than only indirectly
deduced~\cite{kim25, kang25, bohm26}.

In conventional superconductors with dispersive bands, characteristic
length scales such as the Cooper-pair size and the coherence length 
are set by the band curvature and the pairing gap. Although these 
quantities are physically distinct, within weak-coupling BCS theory 
they scale identically and are controlled by the ratio of the Fermi 
velocity $v_\mathrm{F}$ to the superconducting gap $\Delta_0$, 
namely 
$
\xi_{\text{pair}} \sim \xi_{\text{BCS}} \sim \hbar v_\mathrm{F}/\Delta_0
$ 
in the BCS limit~\cite{deGennes66, annett04, leggett}. 
Since $\Delta_0$ is exponentially small in this 
regime, the coherence length becomes very large and Cooper pairs 
strongly overlap. By contrast, in flat-band systems the Fermi 
velocity is ill defined due to the absence of band dispersion, 
and conventional BCS length scales consequently lose their
meaning, motivating geometry-based characterizations of 
superconducting correlations~\cite{tian23}.

Recently, several distinct superconducting length scales have 
been investigated in flat-band 
systems~\cite{iskin23, hu23, thumin24, iskin24c, iskin25, li25, 
virtanen25, lee25, xiao25, oh25}. 
For instance, it has been proposed that the coherence length 
$\xi$ acquires an anomalous quantum-geometric contribution and 
can be expressed as
$
\xi=\sqrt{\xi_{\text{BCS}}^{2}+l_{\text{qm}}^{2}},
$
where $l_{\text{qm}}$ is controlled by the quantum metric~\cite{hu23}. 
Within this picture, $\xi$ remains finite in the flat-band 
limit and is bounded from below by $l_{\text{qm}}$. By contrast, 
an alternative approach based on extracting characteristic length 
scales from the spatial decay of correlation functions within 
real-space Bogoliubov-de Gennes frameworks~\cite{thumin24} has reached 
the opposite conclusion, namely that $\xi$ is disconnected from the 
quantum metric and exhibits no lower bound~\cite{scirep}. 
Furthermore, quantum-geometric effects on the Ginzburg-Landau 
coherence length, as well as on the sizes of two-body bound states 
and Cooper pairs, have been examined using BCS-BEC crossover and 
localization-tensor formalisms~\cite{iskin24c, iskin25}. These 
studies indicate that the various superconducting length scales 
depend on the quantum metric in more intricate and mutually distinct 
ways, suggesting that their relation to quantum geometry is not 
described by a single, unified behavior.

In this work, we analyze characteristic superconducting length scales 
in multiband systems with perfectly flat bands under the assumptions 
of uniform pairing and the presence of time-reversal and 
sublattice-exchange symmetries. 
These assumptions allow us to obtain analytically-tractable 
expressions and, more importantly, to isolate the role of quantum 
geometry by suppressing conventional dispersive contributions and 
enforcing pairing at zero center-of-mass momentum. 
Focusing on the size of the 
lowest-lying two-body bound state, the average Cooper-pair size within 
mean-field BCS-BEC crossover theory, and the zero-temperature 
coherence length obtained from Gaussian fluctuations, we show that 
the pair size remains finite and small in the weak-coupling limit 
and is governed by the quantum geometry of the Bloch states. 
By contrast, the coherence length exhibits qualitatively different 
behavior and diverges in the dilute and insulating limits. 
These results are consistent with and complementary to earlier 
studies emphasizing quantum-geometric contributions to 
superconducting coherence, while clarifying that pair size and 
coherence length are distinct physical quantities in flat-band 
superconductors. More importantly, by explicitly analyzing the 
Creutz and $\chi$ lattices considered in Ref.~\cite{thumin24}, 
our findings help reconcile recent disagreements regarding 
superconducting length scales in flat-band 
systems~\cite{hu23, thumin24, iskin25, scirep}.

The central role of the quantum metric in determining 
the characteristic length scales can be understood from the fact 
that, in perfectly flat bands, conventional kinetic energy is 
quenched and does not provide a length scale for particle motion. 
In this regime, the spatial structure of paired states is governed 
instead by the geometry of the Bloch wave functions. The quantum 
metric quantifies how rapidly Bloch states change in momentum space, 
and through Fourier transformation this directly controls the spatial 
spread of the corresponding wave functions. As a result, it sets the 
intrinsic size of bound states and Cooper pairs. More generally, 
since all interband processes involve overlaps between Bloch states, 
the quantum metric determines the effective mobility of pairs and 
therefore also governs the associated length scales. In this sense, 
quantum geometry replaces band dispersion as the fundamental quantity 
controlling superconducting correlations in flat-band superconductors. 

The remainder of this paper is organized as follows. 
In Sec.~\ref{sec:tF}, we introduce the multiband Hubbard model 
in reciprocal space and define the characteristic superconducting 
length scales considered in this work. In Sec.~\ref{section:numerics}, 
we present numerical results for two representative all-flat-band 
lattices and discuss their implications in the context of the 
recent controversy. Finally, Sec.~\ref{sec:conc} summarizes our 
main findings and conclusions.

\section{Theoretical Formalism}
\label{sec:tF}

In our theoretical formulation, we start from a tight-binding Hubbard 
model defined on a lattice with a multi-orbital basis and transform it 
to reciprocal space. For simplicity, we assume time-reversal symmetry 
and spatially uniform pairing, and restrict our attention to two-band 
Hubbard models with perfectly flat bands. This setting allows us to 
cleanly isolate and analyze the impact of quantum geometry on the 
resulting characteristic length scales, free from conventional 
contributions associated with band dispersion.

\subsection{Multiband Hubbard Hamiltonian}
\label{sec:mHH}

In general, the multi-orbital Hubbard Hamiltonian can be written as 
\begin{align}
\mathcal{H} = \sum_\sigma \mathcal{H}_\sigma + \mathcal{H}_{\uparrow \downarrow}.
\end{align} 
Here, the single-particle contribution
$
\mathcal{H}_\sigma = -\sum_{\mathrm{i}\mathrm{i}'SS'} t^\sigma_{\mathrm{i}S, \mathrm{i}'S'} 
c^\dagger_{S \mathrm{i} \sigma} c_{S' \mathrm{i}' \sigma}
$
describes hopping between lattice sites, where 
$c^\dagger_{S \mathrm{i} \sigma}$ creates a particle with spin 
$\sigma = \{\uparrow, \downarrow\}$ on the sublattice site 
$S = \{A, B\}$ within the unit cell $\mathrm{i}$, and 
$t^\sigma_{\mathrm{i} S, \mathrm{i}' S'}$ denotes the hopping amplitude 
between  sublattice site $S'$ in unit cell $\mathrm{i}'$  and sublattice site $S$ in unit cell $\mathrm{i}$. 
We assume short-ranged interparticle interactions described by
$
\mathcal{H}_{\uparrow \downarrow} = - U \sum_{\mathrm{i} S} 
c^\dagger_{\mathrm{i} S \uparrow} c^\dagger_{\mathrm{i} S \downarrow} 
c_{\mathrm{i} S \downarrow} c_{\mathrm{i} S \uparrow},
$
which corresponds to a local attraction of strength $U \ge 0$ between 
opposite-spin particles occupying the same site.

To transform the Hamiltonian to reciprocal space, we use the Fourier 
expansion
$
c^\dagger_{S \mathrm{i} \sigma} = \frac{1}{\sqrt{N_c}}\sum_{\mathbf{k}} 
e^{-i \mathbf{k} \cdot \mathbf{r}_{\mathrm{i} S}} 
c^\dagger_{S \mathbf{k} \sigma},
$
where $N_c$ is the number of unit cells, crystal momentum $\mathbf{k}$ 
runs over the first Brillouin zone (BZ) satisfying
$
\sum_{\mathbf{k} \in \mathrm{BZ}} 1 = N_c,
$
and $\mathbf{r}_{\mathrm{i} S}$ denotes the position of 
sublattice site $S$ in unit cell $\mathrm{i}$. 
Since each unit cell contains two sublattice sites, 
the total number of lattice sites is $N = 2 N_c$.
Under this transformation, the non-interacting term becomes
$
\mathcal{H}_\sigma = \sum_{\mathbf{k}} h^\sigma_{S S' \mathbf{k}} 
c^\dagger_{S \mathbf{k} \sigma} c_{S' \mathbf{k} \sigma},
$
where
$
h^\sigma_{S S' \mathbf{k}} = -\frac{1}{N_c} 
\sum_{\mathrm{i} \mathrm{i}'} 
t^\sigma_{\mathrm{i} S,\mathrm{i}' S'} 
e^{i \mathbf{k} \cdot 
(\mathbf{r}_{\mathrm{i} S}-\mathbf{r}_{\mathrm{i}'S'})}
$
defines the $2 \times 2$ Bloch Hamiltonian in the sublattice basis. 
This leads to the eigenvalue problem
\begin{align}
\sum_{S'} h^\sigma_{S S' \mathbf{k}} n_{S' \mathbf{k} \sigma} 
= \varepsilon_{n \mathbf{k} \sigma} n_{S \mathbf{k} \sigma},
\label{eqn:hbloch}
\end{align}
where $\varepsilon_{n \mathbf{k} \sigma}$ with $n = \{1,2\}$ is the 
energy of the $n$th Bloch band and 
$
n_{S \mathbf{k} \sigma} = \langle S | n_{\mathbf{k} \sigma} \rangle
$ 
is the sublattice projection of the periodic part of the corresponding 
Bloch state $| n_{\mathbf{k} \sigma} \rangle$. Transforming further 
to the band basis,
$
c^\dagger_{n \mathbf{k} \sigma} = \sum_S n_{S \mathbf{k} \sigma} 
c^\dagger_{S \mathbf{k} \sigma},
$
the non-interacting Hamiltonian takes the diagonal form
\begin{align}
\mathcal{H}_\sigma = \sum_{n\mathbf{k}} \varepsilon_{n \mathbf{k} \sigma} 
c^\dagger_{n \mathbf{k} \sigma} c_{n \mathbf{k} \sigma}.
\label{eqn:Hsigma}
\end{align}
We focus on systems with two perfectly flat bands that possess 
time-reversal and particle-hole symmetries, implying
$
\varepsilon_{n,-\mathbf{k},\downarrow} = \varepsilon_{n \mathbf{k} \uparrow} 
\equiv \varepsilon_n,
$
with $\varepsilon_2 = -\varepsilon_1 \equiv \epsilon$, and
$
n^*_{S,-\mathbf{k},\downarrow} = n_{S \mathbf{k}\uparrow} 
\equiv n_{S\mathbf{k}}.
$

In reciprocal space, the onsite multi-orbital Hubbard 
interaction becomes
$
\mathcal{H}_{\uparrow \downarrow} = -\frac{U}{N_c} 
\sum_{S \mathbf{k} \mathbf{k}' \mathbf{q}} 
c^\dagger_{S, \mathbf{k}+\frac{\mathbf{q}}{2}, \uparrow} 
c^\dagger_{S, -\mathbf{k}+\frac{\mathbf{q}}{2}, \downarrow} 
c_{S, -\mathbf{k}'+\frac{\mathbf{q}}{2}, \downarrow} 
c_{S, \mathbf{k}'+\frac{\mathbf{q}}{2}, \uparrow},
$
and, upon transforming to the band basis, takes the form
\begin{align}
\mathcal{H}_{\uparrow\downarrow} = \frac{1}{N_c} 
\sum_{\substack{n m n' m' \\ \mathbf{k}\mathbf{k}'\mathbf{q}}}
U_{n'm'\mathbf{k}'}^{nm\mathbf{k}}(\mathbf{q})
b_{nm}^\dagger(\mathbf{k}, \mathbf{q})
b_{n'm'}(\mathbf{k}', \mathbf{q}).
\label{eqn:Hud}
\end{align}
Here, the interaction matrix elements
$
U_{n'm'\mathbf{k}'}^{nm\mathbf{k}}(\mathbf{q}) = -U\sum_S 
n^*_{S, \mathbf{k}+\frac{\mathbf{q}}{2}, \uparrow}
m^*_{S, -\mathbf{k}+\frac{\mathbf{q}}{2}, \downarrow}
{m'}_{S, -\mathbf{k}'+\frac{\mathbf{q}}{2}, \downarrow}
{n'}_{S, \mathbf{k}'+\frac{\mathbf{q}}{2}, \uparrow}
$
are nontrivially dressed by the Bloch factors, and
$
b_{nm}^\dagger (\mathbf{k}, \mathbf{q}) =
c^\dagger_{n,\mathbf{k}+\frac{\mathbf{q}}{2}, \uparrow}
c^\dagger_{m,-\mathbf{k}+\frac{\mathbf{q}}{2}, \downarrow}
$
creates a pair of opposite-spin particles in bands $n$ and $m$ with 
relative momentum $\mathbf{k}$ and total momentum $\mathbf{q}$. 
Equations~(\ref{eqn:Hsigma}) and~(\ref{eqn:Hud}) together provide an 
exact reciprocal-space representation of the multi-orbital Hubbard 
model, which we refer to as the multiband Hubbard model throughout 
this paper.

Up to this point, the formulation is general and 
does not rely on any specific assumptions about the band structure 
or pairing. In the following, we specialize to perfectly flat bands 
and impose uniform pairing, which allow us to obtain simplified 
expressions.

\subsection{Pair size for the two-body problem}
\label{section:twobody}

To gain direct insight into the quantum-geometric 
origin of correlation lengths in flat-band superconductors, we consider 
the two-body problem associated with the multiband Hubbard model 
introduced in Eqs.~(\ref{eqn:hbloch}) -~(\ref{eqn:Hud}), consisting 
of one spin-$\uparrow$ and one spin-$\downarrow$ particle on the 
lattice~\cite{iskin21}. 
The bound states of this problem can be obtained exactly by 
employing the general ansatz
$
|\Psi_\mathbf{q} \rangle = \sum_{nm\mathbf{k}} 
\alpha^\mathbf{q}_{nm\mathbf{k}} 
c^\dagger_{n,\mathbf{k} + \frac{\mathbf{q}}{2},\uparrow} 
c^\dagger_{m,-\mathbf{k} + \frac{\mathbf{q}}{2},\downarrow} 
| 0 \rangle,
$
where $| 0 \rangle$ denotes the vacuum state. Here, $\mathbf{q}$ is the 
center-of-mass momentum of the spin-singlet pair, and the complex 
variational parameters satisfy the fermionic exchange symmetry
$
\alpha^\mathbf{q}_{nm\mathbf{k}} = \alpha^\mathbf{q}_{mn,-\mathbf{k}},
$
together with the normalization condition
$
\sum_{nm\mathbf{k}} |\alpha^\mathbf{q}_{nm\mathbf{k}}|^2 = 1.
$
Spin-triplet bound states are excluded due to the onsite nature of the 
Hubbard interaction.

The energy $E_\mathbf{q}$ of the two-body continuum and bound states is
determined by minimizing
$
\langle \Psi_{\mathbf{q}} | \mathcal{H} - E_\mathbf{q}| \Psi_\mathbf{q} \rangle
$
with respect to $\alpha^\mathbf{q}_{nm\mathbf{k}}$, which leads to the
linear equations
$
(\varepsilon_{n} + \varepsilon_{m} - E_{\mathbf{q}} )
\alpha^\mathbf{q}_{nm\mathbf{k}} =
\frac{U}{N_c} \sum_{S} \beta_{S\mathbf{q}}
n^*_{S,\mathbf{k}+\frac{\mathbf{q}}{2}}
m_{S,\mathbf{k}-\frac{\mathbf{q}}{2}}.
$
Here, $E_\mathbf{q}$ plays the role of a Lagrange multiplier enforcing the
normalization condition.
Assuming time-reversal symmetry, we have introduced the dressed 
parameters
$
\beta_{S\mathbf{q}} = \sum_{nm\mathbf{k}} \alpha^{\mathbf{q}}_{nm\mathbf{k}} 
n_{S,\mathbf{k}+\frac{\mathbf{q}}{2}}
m^*_{S,\mathbf{k}-\frac{\mathbf{q}}{2}},
$
whose nonzero values serve as order parameters for the two-body bound 
states. For a given $\mathbf{q}$, the allowed values of $E_\mathbf{q}$ 
can therefore be obtained by solving an eigenvalue problem of dimension 
$4N_c \times 4N_c$.

Alternatively, the bound-state energies can be determined from the 
nonlinear eigenvalue problem~\cite{iskin21}
$
\mathbf{G}^\mathbf{q} \pmb{\beta}_\mathbf{q} = \mathbf{0},
$
where
\begin{align}
G^\mathbf{q}_{SS'} = \delta_{SS'} - \frac{U}{N_c} \sum_{nm\mathbf{k}} 
\frac{n_{S,\mathbf{k}+\frac{\mathbf{q}}{2}} 
m^*_{S,\mathbf{k}-\frac{\mathbf{q}}{2}} 
n^*_{S',\mathbf{k}+\frac{\mathbf{q}}{2}} 
m_{S',\mathbf{k}-\frac{\mathbf{q}}{2}}}
{\varepsilon_{n} + \varepsilon_{m} - E_\mathbf{q} }
\label{eqn:G0}
\end{align}
is a $2 \times 2$ Hermitian matrix. Here, $\delta_{SS'}$ is the Kronecker 
delta, and nontrivial solutions for $\pmb{\beta}_\mathbf{q}$ correspond 
to values of $E_\mathbf{q}$ at which $\mathbf{G}_\mathbf{q}$ develops a 
zero eigenvalue.
We focus on lattices that satisfy the uniform-pairing condition, namely 
that the eigenvector of $\mathbf{G}_\mathbf{q}$ associated with the 
lowest-energy solution has equal weight on the two sublattices. In the 
$\mathbf{q} \to \mathbf{0}$ limit, this implies
$
\beta_{A\mathbf{q}} = \beta_{B\mathbf{q}} \equiv \beta_\mathbf{q}.
$
We have verified numerically that this condition holds for the 
lowest-lying two-body bound states of the flat-band lattices considered 
in this work, where sublattice-exchange symmetry is present.

The size of a two-body bound state is characterized by the trace of the 
two-body localization tensor~\cite{iskin25}, 
whose matrix elements are defined as
\begin{equation}
(\xi^2_{2b})_{ij} = \sum_{\mathrm{i}S \mathrm{i}'S'} 
\bar{r}_i \bar{r}_j \,
| \psi^\mathbf{q}_{SS'}(\bar{\mathbf{r}})|^2,
\end{equation}
where $\bar{r}_i$ ($i=\{x,y,z\}$) denotes the components of the relative 
coordinate
$
\bar{\mathbf{r}} = \mathsf{r}_{\mathrm{i} S} - \mathsf{r}_{\mathrm{i}' S'},
$
with  $\mathsf{r}_{\mathrm{i} S}$ and $\mathsf{r}_{\mathrm{i}' S'}$ 
denoting the positions of lattice sites $(\mathrm{i},S)$ and $(\mathrm{i}',S')$,
respectively. The spin degrees of freedom are carried separately by the operators, 
and the relative coordinates correspond to the separation between two 
particles forming a pair.
The relative wave function is given by
$
\psi^\mathbf{q}_{SS'}(\bar{\mathbf{r}}) = \frac{1}{N_c} 
\sum_{nm\mathbf{k}} 
e^{i \mathbf{k} \cdot \bar{\mathbf{r}}} 
\alpha^\mathbf{q}_{nm\mathbf{k}} 
n_{S,\mathbf{k}+\frac{\mathbf{q}}{2}} 
m_{S',\mathbf{k}-\frac{\mathbf{q}}{2}},
$
up to an overall plane-wave factor associated with the center-of-mass 
motion.
In the following, we restrict attention to the lowest two-body bound 
state at $\mathbf{q}=\mathbf{0}$, for which
$
\alpha^\mathbf{0}_{nm\mathbf{k}} =
\frac{U |\beta_\mathbf{0}| \delta_{nm}}
{N_c (2\varepsilon_n - E_b)}
$
is independent of $\mathbf{k}$. The corresponding bound-state energy 
$E_\mathbf{0} \equiv E_b$ is
$
E_b = -\frac{U + \sqrt{U^2 + 16\epsilon^2}}{2},
$
which is valid for all $U$ and follows from the condition
$
\sum_{SS'} G^\mathbf{0}_{SS'} = 0,
$
or equivalently,
$
1 = \frac{U}{2} \sum_n \frac{1}{2\varepsilon_n - E_b},
$
where we select the solution satisfying binding energy $|E_b| \to U$ 
in the strong-coupling limit $U/t \gg 1$.

Using the identity
$ 
\bar r_i e^{-i\mathbf{k}\cdot\bar{\mathbf r}}
= i\partial_{k_i}e^{-i\mathbf{k}\cdot\bar{\mathbf r}},
$ 
multiplication by the relative coordinate in real space can be 
converted into a momentum derivative acting on the plane-wave factor. 
Consequently, the pair size is governed by derivatives of the two-body 
wave function in reciprocal space. In particular, 
using integration by parts, the localization tensor can be expressed as
$
(\xi^2_{2b})_{ij} = \sum_{nmSS'\mathbf{k}} 
\alpha^\mathbf{0}_{nn\mathbf{k}} 
(\alpha^\mathbf{0}_{mm\mathbf{k}})^* 
\partial_{k_i} ( n_{S \mathbf{k}} n^*_{S' \mathbf{k}} )
\partial_{k_j} ( m^*_{S \mathbf{k}} m_{S' \mathbf{k}} ),
$
which yields a purely interband contribution,
$
(\xi^2_{2b})_{ij} \equiv (\xi^2_{2b})^{\mathrm{inter}}_{ij},
$
with
\begin{align}
(\xi^2_{2b})^{\mathrm{inter}}_{ij} =
\frac{8\epsilon^2}{4\epsilon^2 + E_b^2}
\frac{1}{N_c} \sum_{\mathbf{k}} g^\mathbf{k}_{ij}.
\label{eqn:xi2b}
\end{align}
For the two-band models considered here, the quantum-metric tensors of 
the two flat bands are identical,
$
g^{1\mathbf{k}}_{ij} = g^{2\mathbf{k}}_{ij} \equiv g^\mathbf{k}_{ij},
$
where
\begin{align}
g^\mathbf{k}_{ij} = 2\,\mathrm{Re}\,
\langle \partial_{k_i} 1_\mathbf{k} | 2_\mathbf{k} \rangle
\langle 2_\mathbf{k} | \partial_{k_j} 1_\mathbf{k} \rangle.
\label{eqn:qm}
\end{align}
More generally, the quantum-metric tensor of the $n$th Bloch band is
$
g^{n\mathbf{k}}_{ij} = 2\,\mathrm{Re} \sum_{m \neq n}
\langle \partial_{k_i} n_\mathbf{k} | m_\mathbf{k} \rangle
\langle m_\mathbf{k} | \partial_{k_j} n_\mathbf{k} \rangle,
$
which is, by construction, a real symmetric 
matrix~\cite{Provost80, resta11}. 
Consequently, the size of the lowest two-body bound state is entirely
governed by the quantum metric of the flat bands. Importantly, 
Eq.~(\ref{eqn:xi2b}) is exact for the $\mathbf{q}=\mathbf{0}$ bound state 
for all values of $U$.

Finally, we note that the effective-mass tensor of the lowest two-body 
bound state can be extracted from Eq.~(\ref{eqn:G0}) by expanding
$
E_\mathbf{q} = E_b + \frac{1}{2}\sum_{ij} (M^{-1}_{2b})_{ij} q_i q_j + \cdots 
$
in the $\mathbf{q} \to \mathbf{0}$ limit where $i = \{x,y,z\}$~\cite{iskin21}. 
This yields a purely interband contribution,
$
(M^{-1}_{2b})_{ij} \equiv (M^{-1}_{2b})^{\mathrm{inter}}_{ij},
$
with
\begin{align}
(M^{-1}_{2b})^{\mathrm{inter}}_{ij}
= \frac{4\epsilon^2(4\epsilon^2 - E_b^2)}{E_b (4\epsilon^2 + E_b^2)} 
\frac{1}{N_c} \sum_{\mathbf{k}} g^\mathbf{k}_{ij},
\label{eqn:Mij}
\end{align}
which is exact for the two-body problem. 
Note that $(M^{-1}_{2b})_{ij} \to 0$ as $E_b \to -2\epsilon$ when $U = 0$.
This behavior originates from the perfectly flat single-particle
bands, for which the constituent particle masses diverge, resulting
in a divergent effective two-body mass. By contrast, 
$
(\xi^2_{2b})_{ij} \to 
\frac{1}{N_c} \sum_{\mathbf{k}} g^\mathbf{k}_{ij}
$ 
remains finite in the same limit, since it is controlled by the quantum 
geometry of the flat-band Bloch states rather than by band dispersion.
Moreover, this expression provides the upper bound for the pair size.
In dilute flat-band superconductors, the effective-mass tensor of Cooper 
pairs is well approximated by the same expression, with small corrections 
appearing at higher fillings~\cite{iskin24c}. 
We therefore employ Eq.~(\ref{eqn:Mij}) in 
Sec.~\ref{section:numerics} to analyze the coherence length in the 
many-body problem.

\subsection{Pair size for the many-body problem}
\label{section:manybody}

We now turn to the average size of Cooper pairs within the mean-field 
BCS theory at zero temperature ($T=0$), under the same assumptions as in 
the two-body problem, namely time-reversal symmetry and the 
uniform-pairing condition~\cite{iskin25}. 
Under these conditions, the onsite 
mean-field order parameter is independent of the sublattice and can be 
written as
$
\Delta_A = \Delta_B \equiv \Delta_0,
$
which can be chosen to be a positive real number without loss of 
generality. For the two-band Hubbard models with perfectly flat bands 
considered in this paper, the corresponding BCS ground state is
$
|\text{BCS}\rangle = \prod_{n\mathbf{k}}
( u_n + v_n \,
c^\dagger_{n\mathbf{k}\uparrow}
c^\dagger_{n,-\mathbf{k}\downarrow})
|0\rangle,
$
where the coherence factors are
$
u_n = \sqrt{\frac{1}{2} + \frac{\xi_n}{2E_n}},
$
and
$
v_n = \sqrt{\frac{1}{2} - \frac{\xi_n}{2E_n}}.
$
Here, $\xi_n = \varepsilon_n - \mu$ denotes the band energy measured from 
the chemical potential $\mu$, and
$
E_n = \sqrt{\xi_n^2 + \Delta_0^2}
$
is the quasiparticle spectrum. The parameters $\Delta_0$ and $\mu$ are 
determined self-consistently from the mean-field gap and number 
equations,
\begin{align}
1 &= \frac{U}{2} \sum_n \frac{1}{2E_n},
\label{eqn:gap}
\\
F &= 1 - \frac{1}{2} \sum_n \frac{\xi_n}{E_n},
\label{eqn:mu}
\end{align}
where the filling $0 \le F \le 2$ denotes the total number of particles 
per lattice site. Equations~(\ref{eqn:gap}) and~(\ref{eqn:mu}) are 
well-established for providing a qualitatively accurate description 
of the BCS-BEC crossover physics at $T = 0$~\cite{strinati18}.

We define the properly normalized Cooper-pair wave function 
as~\cite{deGennes66, iskin25}
\begin{equation}
\Phi(\mathsf{r}_{\mathrm{i} S}, \mathsf{r}_{\mathrm{i}' S'}) =
\frac{1}{A_\mathrm{Cp}}
\langle \text{BCS} |
\psi^\dagger_\uparrow(\mathsf{r}_{\mathrm{i} S})
\psi^\dagger_\downarrow(\mathsf{r}_{\mathrm{i}' S'})
| \text{BCS} \rangle,
\label{eqn:phi}
\end{equation}
and expand the field operators in the Bloch basis,
$
\psi^\dagger_\sigma(\mathsf{r}_{\mathrm{i} S}) =
\sum_{n\mathbf{k}}
\phi^*_{n\mathbf{k}\sigma}(\mathsf{r}_{\mathrm{i} S})
c^\dagger_{n\mathbf{k}\sigma},
$
where
$
\phi_{n\mathbf{k}\sigma}(\mathsf{r}_{\mathrm{i} S}) =
\langle \mathsf{r}_{\mathrm{i} S} | n\mathbf{k}\sigma \rangle =
\frac{e^{i\mathbf{k}\cdot\mathsf{r}_{\mathrm{i} S}}}{\sqrt{N_c}}
n_{S\mathbf{k}\sigma}
$
is the Bloch wave function. This yields
$
\Phi(\mathsf{r}_{\mathrm{i} S}, \mathsf{r}_{\mathrm{i}' S'}) =
\frac{1}{A_\mathrm{Cp} N_c}
\sum_{n\mathbf{k}}
e^{-i\mathbf{k}\cdot\bar{\mathbf{r}}}
u_n v_n \,
n^*_{S\mathbf{k}} n_{S'\mathbf{k}},
$
where $\bar{\mathbf{r}}=\mathsf{r}_{\mathrm{i} S}-\mathsf{r}_{\mathrm{i}' S'}$
is the relative coordinate. The normalization constant is
$
A_\mathrm{Cp} = \frac{N_c}{2} \sum_n \frac{\Delta_0^2}{2E_n^2},
$
which corresponds to the total number of condensed Cooper 
pairs~\cite{leggett}.

In direct analogy with the two-body problem, the average size of Cooper 
pairs is characterized by the trace of the Cooper-pair localization 
tensor~\cite{iskin25},
\begin{equation}
(\xi^2_\mathrm{Cp})_{ij} =
\sum_{\mathrm{i}S \mathrm{i}'S'}
\bar{r}_i \bar{r}_j
\left|
\Phi(\mathsf{r}_{\mathrm{i} S}, \mathsf{r}_{\mathrm{i}' S'})
\right|^2.
\end{equation}
Converting the relative-coordinate factors into momentum derivatives and
using integration by parts, this expression can be written as
$
(\xi^2_\mathrm{Cp})_{ij} =
\frac{1}{A_\mathrm{Cp}}
\sum_{nmSS'\mathbf{k}}
u_n v_n \, u_m v_m \,
\partial_{k_i}\!\left( n_{S\mathbf{k}} n^*_{S'\mathbf{k}} \right)
\partial_{k_j}\!\left( m^*_{S\mathbf{k}} m_{S'\mathbf{k}} \right).
$

For the flat-band models considered here, this reduces to a purely 
interband contribution,
$
(\xi^2_\mathrm{Cp})_{ij} \equiv (\xi^2_\mathrm{Cp})^{\text{inter}}_{ij},
$
given by
\begin{align}
(\xi^2_\mathrm{Cp})^{\text{inter}}_{ij} =
\frac{(E_1 - E_2)^2}{E_1^2 + E_2^2}
\frac{1}{N_c} \sum_{\mathbf{k}} g^\mathbf{k}_{ij}.
\label{eqn:xiCp}
\end{align}
Thus, as in the two-body case, the average size of Cooper pairs is 
entirely governed by the quantum-metric tensor of the flat bands. 
Equation~(\ref{eqn:xiCp}) is valid for all interaction strengths $U$ and 
fillings $F$ within the zero-temperature mean-field theory. 
Note that $(\xi^2_\mathrm{Cp})_{ij} \to 0$ as $E_1 \to E_2$, i.e., at 
half filling with $\mu = 0$ in particle-hole symmetric systems (provided 
that $\Delta_0 \ne 0$), and at any filling in the $U/t \to \infty$ limit.

\subsection{Zero-temperature coherence length}
\label{section:cl}

To facilitate a direct comparison with the average size of Cooper pairs, 
we next introduce the zero-temperature coherence length for a multiband 
Hubbard model under the same assumptions of time-reversal symmetry and 
the uniform-pairing condition~\cite{iskin24c}. 
Unlike the Cooper-pair size, this is a 
beyond-mean-field quantity that is defined through the effective 
Gaussian action describing fluctuations of the superconducting order 
parameter about the mean-field saddle point.
We assume that, in the long-wavelength limit $Q \to 0$, fluctuations of 
the order parameter remain uniform across the lattice. Accordingly, we 
introduce the Hubbard-Stratonovich field
$
\Delta_{SQ} = \Delta_0 + \Lambda_Q,
$
which is taken to be independent of the sublattice index, with
$
\Lambda_{AQ} = \Lambda_{BQ} \equiv \Lambda_Q
$
describing fluctuations around the saddle-point value $\Delta_0$. Here,
$
Q \equiv (\mathbf{q}, i\nu_l)
$
is a collective index, where $\nu_l = 2\pi l T$ denotes the bosonic 
Matsubara frequency, $T$ is the temperature, and $l = 0, \pm1, \pm2, \ldots$ 
is an integer. Throughout, we use units in which $\hbar = 1$ and 
$k_\mathrm{B} = 1$.

Following the Gaussian fluctuation formalism developed in
Ref.~\cite{iskin24c}, the effective action for order-parameter fluctuations 
can be expressed in terms of the fluctuation matrix
$
\pmb{\mathcal{M}}^Q,
$
which governs amplitude and phase modes. In the present work, 
we adopt this framework and specialize it to the case of perfectly flat 
two-band systems, allowing us to isolate the geometric contributions to 
the coherence length. The Gaussian action for the order-parameter 
fluctuations can then be written as 
$
\mathcal{S}_\mathrm{G} = \frac{N}{2T} \sum_Q
\begin{pmatrix} \Lambda^*_Q & \Lambda_{-Q} \end{pmatrix}
\pmb{\mathcal{M}}^Q
\begin{pmatrix} \Lambda_Q \\ \Lambda^*_{-Q} \end{pmatrix},
$
where the fluctuation matrix $\pmb{\mathcal{M}}^Q$ plays the role of the 
inverse propagator for amplitude and phase fluctuations.
The zero-temperature coherence length is defined by setting 
$i\nu_l = 0$ and focusing on the amplitude-amplitude sector in the 
long-wavelength limit~\cite{pistolesi96, benfatto02}, 
specifically on the combination
$
\mathcal{M}^\mathbf{q}_{11} + \mathcal{M}^\mathbf{q}_{12}.
$
Expanding this quantity for $\mathbf{q} \to \mathbf{0}$ yields
$
\mathcal{M}^\mathbf{q}_{11} + \mathcal{M}^\mathbf{q}_{12}
= A + \sum_{ij} C_{ij} q_i q_j,
$
where $A$ is the static coefficient and
$
C_{ij} \equiv C^{\text{inter}}_{ij}
$
is a purely interband kinetic coefficient. Explicitly, these coefficients 
are given by
\begin{align}
A &= \frac{\Delta_0^2}{4E_1^3} + \frac{\Delta_0^2}{4E_2^3}, 
\label{eqn:A}
\\
C^{\text{inter}}_{ij} &= \bigg(
\frac{\xi_1^2}{8E_1^3} + \frac{\xi_2^2}{8E_2^3}
- \frac{\xi_1 \xi_2 + E_1 E_2 - \Delta_0^2}
{4E_1 E_2 (E_1 + E_2)}
\bigg)
\frac{1}{N_c} \sum_{\mathbf{k}} g^{\mathbf{k}}_{ij},
\label{eqn:Cij}
\end{align}
where the saddle-point parameters $\Delta_0$ and $\mu$ are determined 
self-consistently from Eqs.~(\ref{eqn:gap}) and~(\ref{eqn:mu}). 

Provided that the trace of $C_{ij}$ is positive, the zero-temperature 
coherence length is well defined and has a purely interband origin,
$
(\xi^2_0)_{ij} \equiv (\xi^2_0)^{\text{inter}}_{ij},
$
characterized by~\cite{iskin24c}
\begin{align}
(\xi^2_0)^{\text{inter}}_{ij} = \frac{C^{\text{inter}}_{ij}}{A}.
\label{eqn:xi0}
\end{align}
Thus, in direct analogy with $(\xi^2_{2b})_{ij}$ and 
$(\xi^2_\mathrm{Cp})_{ij}$, the zero-temperature coherence length is 
entirely governed by the quantum-metric tensor of the flat bands. It is 
also referred to as the Higgs-mode correlation length~\cite{xiao25, oh25}. 
While Eq.~(\ref{eqn:xi0}) formally applies for all interaction strengths 
$U$ and fillings $F$ within the mean-field framework, our numerical 
analysis shows that the trace of $C_{ij}$ may become negative near 
quarter filling. In this regime, $(\xi^2_0)_{ij}$ ceases to be physical, as 
discussed in Sec.~\ref{section:numerics}. Similar to 
$(\xi^2_\mathrm{Cp})_{ij}$, we also note that $(\xi^2_0)_{ij} \to 0$ as 
$E_1 \to E_2$ at any filling in the $U/t \to \infty$ limit.

We emphasize that our analytical expressions are derived for two-band 
systems, where the quantum metric involves a single interband contribution. 
In more general multiband systems, even when all bands are perfectly flat, 
the corresponding characteristic length scales receive contributions 
from all interband processes, rendering the formulation more involved. 
Accordingly, while Eqs.~(\ref{eqn:xi2b}), (\ref{eqn:xiCp}), and 
(\ref{eqn:xi0}) take a more complicated form, i.e., incorporating 
quantum-metric tensors associated with all bands as well as the so-called 
band-resolved quantum metrics, they remain fully determined by the 
underlying quantum geometry and are therefore fundamentally geometric 
in nature~\cite{iskin24c,iskin25}.  

 We also note that parts of the formalism presented in this 
section build on results previously developed in 
Refs.~\cite{iskin21, iskin24, iskin24c, iskin25}, where the two-body 
problem, localization tensors, and quantum-geometric contributions in 
multiband systems were introduced in a more general setting. In the 
present work, we adopt these established frameworks and specialize them 
to the case of perfectly flat bands, which allows us to isolate the 
role of quantum geometry and to directly compare distinct superconducting 
length scales within a unified approach. In particular, our focus 
here is not on the development of the formalism itself, but on its 
application to all-flat-band systems and on clarifying the relation 
between the two-body pair size, the many-body Cooper-pair size, and 
the coherence length.

\section{Numerical Results and Discussion}
\label{section:numerics}

\begin{figure}
\includegraphics[width=0.5\textwidth]{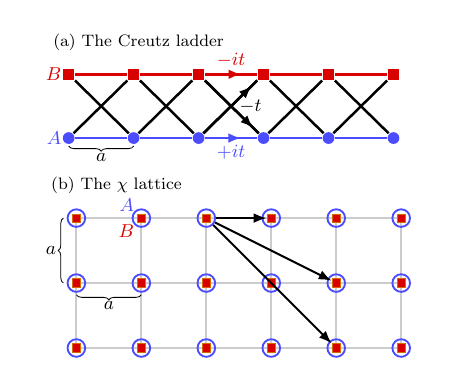}

\caption{\label{fig:lattice}
Illustration of (a) the Creutz ladder and (b) the $\chi$ lattice.
Blue circles (A) and red squares (B) denote the two sublattices, and
the lattice spacing is denoted by $a$.
In the $\chi$ lattice, each lattice site hosts two orbitals (A and B),
and long-range hopping processes connect A and B orbitals on
different lattice sites, as indicated by the arrows.
}
\end{figure}

In the previous section, we introduced the characteristic length
scales of a two-band Hubbard model with perfectly flat bands, assuming
time-reversal symmetry and a uniform-pairing condition. In this
section, we demonstrate these results numerically by analyzing two
representative lattice models: (i) the Creutz ladder and (ii) the
$\chi$ lattice, both of which are depicted in Fig.~\ref{fig:lattice}.
We note that the quantum metric given in Eq.~(\ref{eqn:qm}) 
is gauge invariant under 
$
\mathbf{k}
$-dependent U(1) phase transformations of the Bloch states, 
$ 
| n_\mathbf{k} \rangle \to e^{i \theta_{n\mathbf{k}}} | n_\mathbf{k} \rangle. 
$ 
Since the characteristic length scales introduced in Eqs. 
(\ref{eqn:xi2b}), (\ref{eqn:xiCp}) and (\ref{eqn:xi0}) depend only on 
Brillouin-zone averages of the quantum metric, they are likewise gauge 
invariant which we have also verified this explicitly in our numerical 
calculations.

The Creutz ladder is a one-dimensional, two-legged lattice with two
sublattices. It is described by the Bloch Hamiltonian
$
h_{\downarrow,-k}^* = h_{\uparrow k} \equiv h_k,
$
with~\cite{thumin24}
\begin{align}
h_k =
\begin{bmatrix}
-2t \sin(ka) & -2t \cos(ka) \\
-2t \cos(ka) &  2t \sin(ka)
\end{bmatrix}.
\end{align}
Here, $a$ is the lattice spacing, and $-\pi/a \le k < \pi/a$ defines
the BZ. The Bloch spectrum consists of two perfectly flat bands,
$
\varepsilon_{2k} = -\varepsilon_{1k} = 2t,
$
and the associated Bloch states have sublattice components
$
1_{Ak} = -2_{Bk} = \sin \gamma_k
$
and
$
1_{Bk} = 2_{Ak} = \cos \gamma_k,
$
where $\gamma_k = \pi/4 + ka/2$.
The quantum metric associated with these flat bands is momentum
independent and takes the diagonal form
$
g^k_{ij} = g_k \delta_{ij},
$
with
$
g_k = a^2/2.
$
We would like to note that the assumption of a superconducting 
ground state in the Creutz lattice is not merely a heuristic choice 
but is rigorously supported by unbiased numerical evidence. 
Specifically, numerically exact density matrix renormalization group (DMRG) 
calculations consistently reveal a power-law (algebraic) decay of the pair 
correlation function, i.e., the definitive signature of a superconducting 
phase in one-dimensional systems. 
Furthermore, it has been demonstrated that a refined multi-band mean-field 
approach achieves excellent agreement with exact DMRG results across a 
wide range of coupling strengths, justifying the use of the mean-field 
ansatz in investigating the system's superconducting 
properties~\cite{mondaini18, chan22}.

The $\chi$ lattice is a two-dimensional square lattice with two
sublattices, where the parameter $\chi$ controls the strength of the
long-range hopping process. It is described by the Bloch Hamiltonian
$
h_{\downarrow,-\mathbf{k}}^* = h_{\uparrow \mathbf{k}}
\equiv h_{\mathbf{k}},
$
with~\cite{thumin24, hofmann22, hofmann23}
\begin{align}
h_{\mathbf{k}} =
\begin{bmatrix}
0 & -t e^{-i \gamma_{\mathbf{k}}} \\
-t e^{i \gamma_{\mathbf{k}}} & 0
\end{bmatrix}.
\end{align}
Here, $a$ is the lattice spacing, and
$-\pi/a \le k_x < \pi/a$ and $-\pi/a \le k_y < \pi/a$ define the BZ, with
$
\gamma_{\mathbf{k}} = \chi \big[ \cos(k_x a) + \cos(k_y a) \big].
$
Its Bloch spectrum again consists of two perfectly flat bands,
$
\varepsilon_{2\mathbf{k}} = -\varepsilon_{1\mathbf{k}} = t,
$
and the associated Bloch states have sublattice components
$
1_{A\mathbf{k}} = 2_{B\mathbf{k}} = \frac{1}{\sqrt{2}}
$
and
$
1_{B\mathbf{k}} = -2_{A\mathbf{k}}^*
= \frac{e^{i\gamma_{\mathbf{k}}}}{\sqrt{2}}.
$
The quantum metric associated with these flat bands is
$
g^{\mathbf{k}}_{ij}
= \frac{a^2 \chi^2}{2} \sin(k_i a)\sin(k_j a).
$
In our numerical calculations, we set $\chi = 1$. 
The assumption of a superconducting ground state in the $\chi$ lattice 
is also supported by numerically exact determinant quantum 
Monte Carlo simulations~\cite{hofmann23}.

First, we note that both lattices possess not only time-reversal symmetry
by construction but also sublattice-exchange and particle-hole symmetries.
As a consequence, Eq.~(\ref{eqn:G0}) implies that
$
G^\mathbf{0}_{AA} = G^\mathbf{0}_{BB}
$
and
$
G^\mathbf{0}_{AB} = G^\mathbf{0}_{BA}
$
for the lowest bound ($\mathbf{q} = \mathbf{0}$) two-body state.
These relations imply that the vector
$
\frac{1}{\sqrt{2}} \begin{pmatrix} 1 & \pm 1\end{pmatrix}^\mathrm{T},
$
with $\mathrm{T}$ denoting the transpose, is an eigenvector corresponding
to the eigenvalue
$
G^\mathbf{0}_{AA} \pm G^\mathbf{0}_{AB},
$
respectively. Thus, by setting the $+$ eigenvalue to zero, we can determine
$E_b$ of the lowest bound two-body state. This result
demonstrates that the uniform-pairing condition, i.e., the $+$ eigenvector,
is satisfied exactly for the two-body problem in both lattices, and
therefore that the analyses of the previous sections are applicable.
We have also numerically verified this observation, finding that
$
\beta_{A\mathbf{q}} = \beta_{B\mathbf{q}}
$
holds with numerical exactness as $\mathbf{q} \to \mathbf{0}$. 
Furthermore, the uniform-pairing condition is known to be satisfied 
for the many-body problem in both lattices within the real-space 
Bogoliubov-de Gennes formulation of the mean-field 
theory~\cite{thumin24}. See also Ref.~\cite{mondaini18, chan22}.

\begin{figure*}[htb]
\includegraphics[width=0.25\textwidth,clip=false]{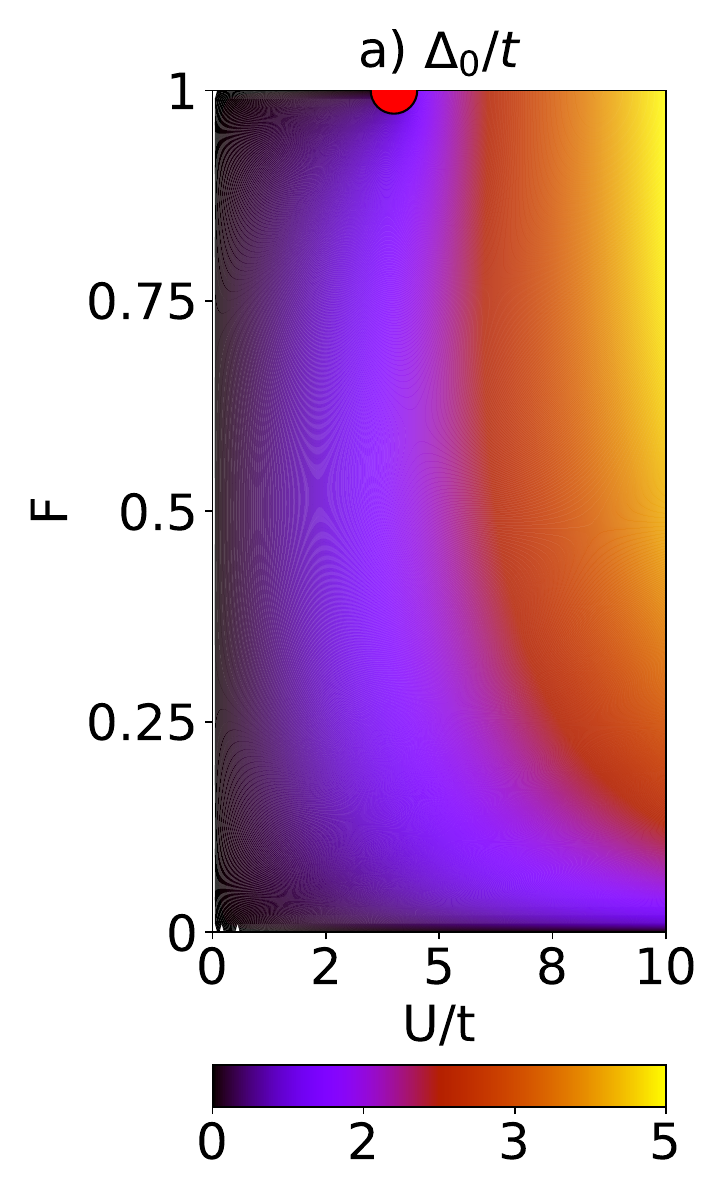}\hfill
\includegraphics[width=0.25\textwidth,clip=false]{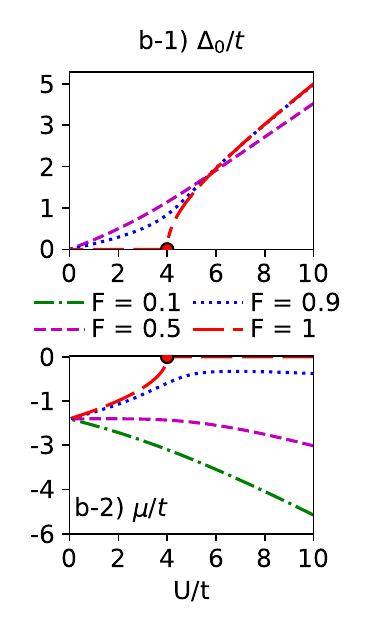}\hfill
\includegraphics[width=0.25\textwidth,clip=false]{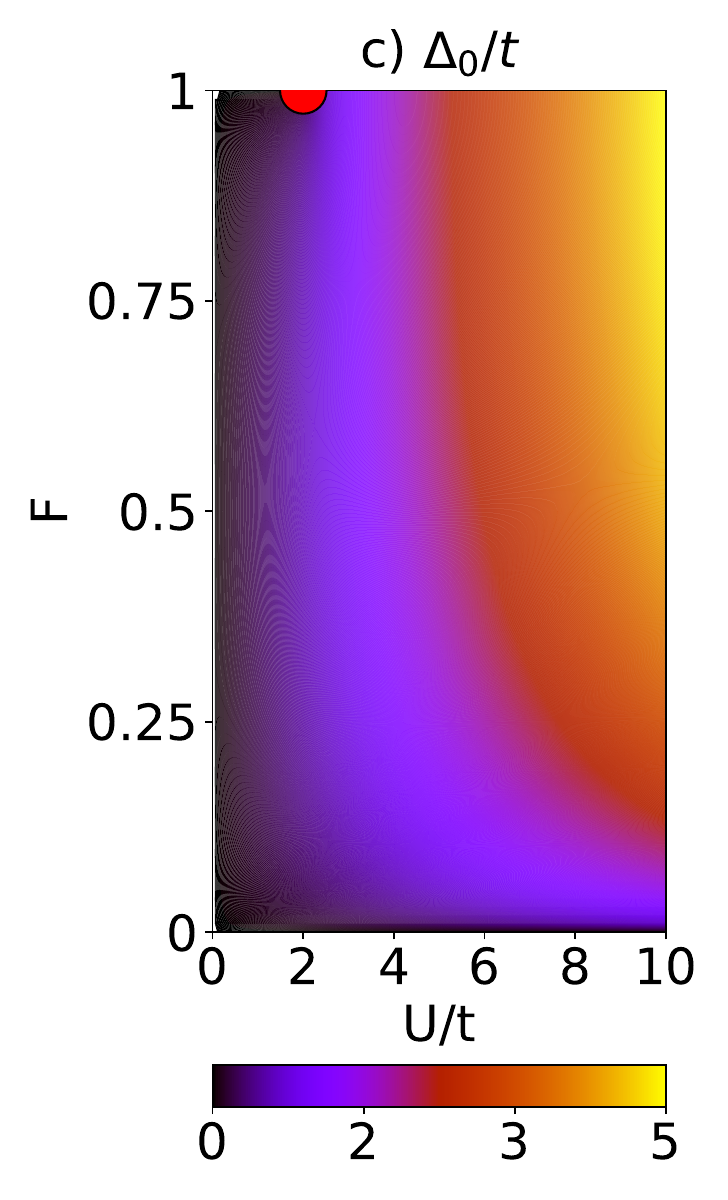}\hfill
\includegraphics[width=0.25\textwidth,clip=false]{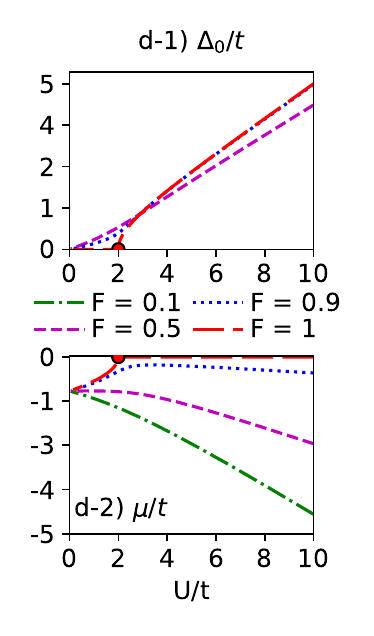}

\caption{\label{fig:gap}
The order parameter $\Delta_0$ is shown in (a) for the Creutz ladder
and in (c) for the $\chi$ lattice as functions of the particle filling
$F$ and the interaction strength $U$.
The corresponding line cuts of 
$\Delta_0$ and the chemical potential $\mu$ are shown in  (b-1) and (b-2) for the Creutz ladder, and in (d-1) and (d-2) for the $\chi$ lattice.  
At half filling ($F = 1$), $\Delta_0$ vanishes and
$\mu$ is pinned within a finite interval, signaling an insulating state,
for interaction strengths below a critical threshold $U < U_c$.
The red dots mark the location of $U_c = 2\epsilon$.
}
\end{figure*}

In Fig.~\ref{fig:gap}, we present the self-consistent solutions
of Eqs.~(\ref{eqn:gap}) and~(\ref{eqn:mu}) for the mean-field
parameters $\Delta_0$ and $\mu$. Owing to particle-hole symmetry,
all results are symmetric about half filling ($F = 1$), and we therefore
present them only in the range $0 \leq F \leq 1$.
For instance, in the $U/t \to 0$ limit, our numerical solutions are
fully consistent with the analytical expectations that
$
\mu = -\epsilon - \frac{U}{4}(1-2F)
$
and
$
\Delta_0 = \frac{U}{2} \sqrt{F(1-F)}
$
for $0 \le F < 1$, and
$
\mu = \epsilon - \frac{U}{4}(3-2F)
$
and
$
\Delta_0 = \frac{U}{2}\sqrt{(F-1)(2-F)}
$
for $1 < F \le 2$. Note that $(-\mu,\Delta_0)$ are solutions of
Eqs.~(\ref{eqn:gap}) and~(\ref{eqn:mu}) for $U$ and $2 - F$,
provided that $(\mu, \Delta_0)$ are solutions for given $U$ and $F$.
In addition, we verified that
$
\mu = - \frac{U}{2}(1-F)
$
and
$
\Delta_0 = \frac{U}{2}\sqrt{F(2-F)}
$
in the $U/t \to \infty$ limit.

The simplicity of the Bloch spectrum further allows an analytic solution
of the mean-field equations at half filling, for which we find
$
\mu = \pm \sqrt{\epsilon^2 - U \epsilon/2}
$
together with $\Delta_0 = 0$ when $U \le U_c = 2\epsilon$, 
and $\mu = 0$ together with
$
\Delta_0 = \sqrt{U^2/4 - \epsilon^2}
$
when $U \ge U_c$. These results are also visible in Fig.~\ref{fig:gap}(b)
for the Creutz ladder and Fig.~\ref{fig:gap}(d) for the $\chi$ lattice.
The critical interaction threshold is given by the band gap, i.e., 
$U_c = 2\epsilon$, and it is marked by a red dot in all panels.
At half filling, when $U < U_c$, the vanishing $\Delta_0$ indicates that
the system remains in the normal state, while the pinning of $\mu$ over
a finite interval signals an insulating phase. The $\pm$ signs in $\mu$
correspond to particle- and hole-like excitations from the band insulator,
respectively, and together they form an insulating dome in the figures.
Previous DMRG and mean-field studies~\cite{mondaini18, chan22} of the Creutz
ladder have already identified the half-filled system as a band insulator,
and our analysis suggests that this insulating state persists up to
$U_c = 2\epsilon$ within the mean-field approximation. For interactions
$U > U_c$, the system transitions to a superconducting phase, which can
account for the finite superfluid weight reported in mean-field
calculations at $U = 8t$~\cite{chan22}. In the remainder of this
discussion, we exclude the parameter regime in which $\Delta_0$ vanishes,
since the length scales introduced in Sec.~\ref{sec:tF} require a nonzero
$\Delta_0$ to begin with.

\begin{figure*}[htb]
\includegraphics[width=0.25\textwidth,clip=false]{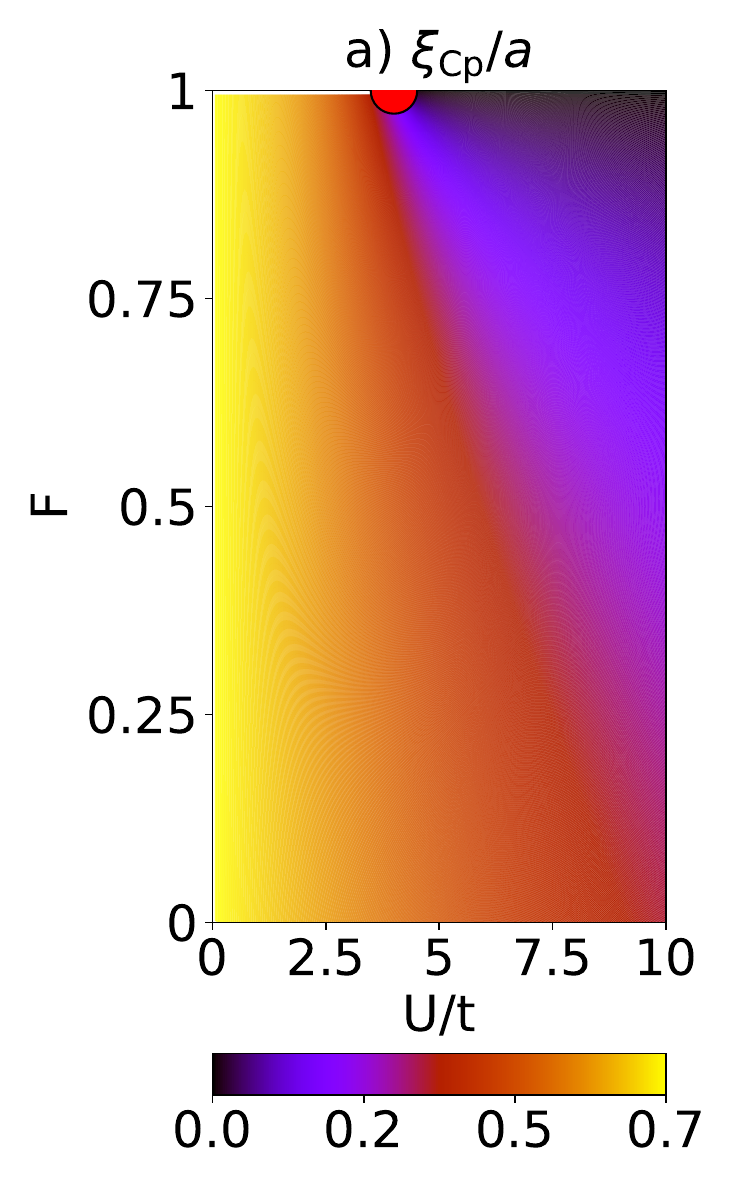}\hfill
\includegraphics[width=0.25\textwidth,clip=false]{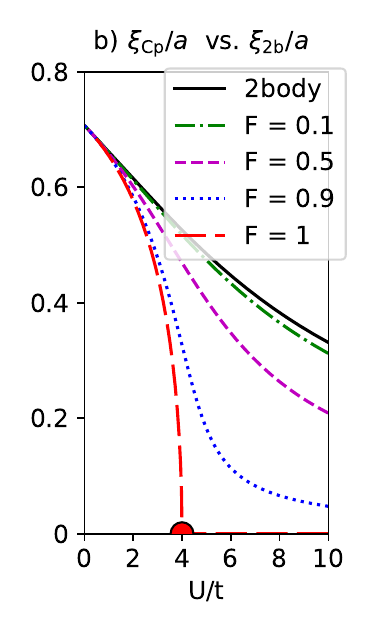}\hfill
\includegraphics[width=0.25\textwidth,clip=false]{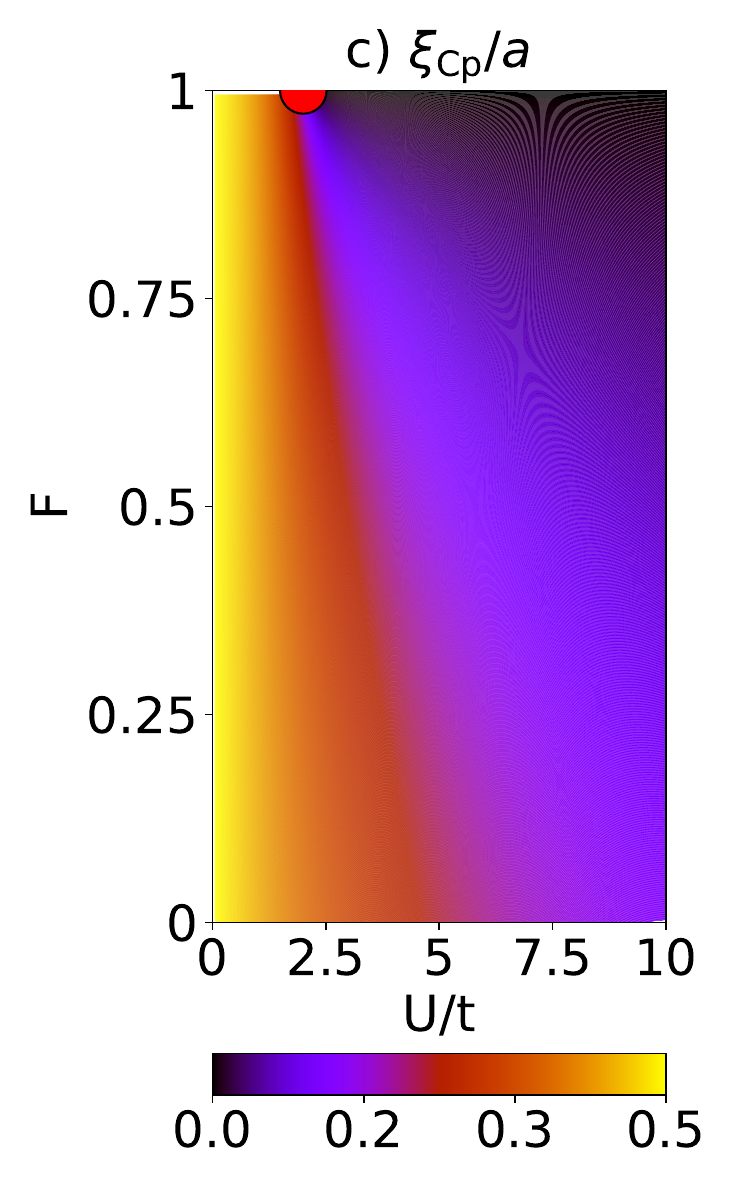}\hfill
\includegraphics[width=0.25\textwidth,clip=false]{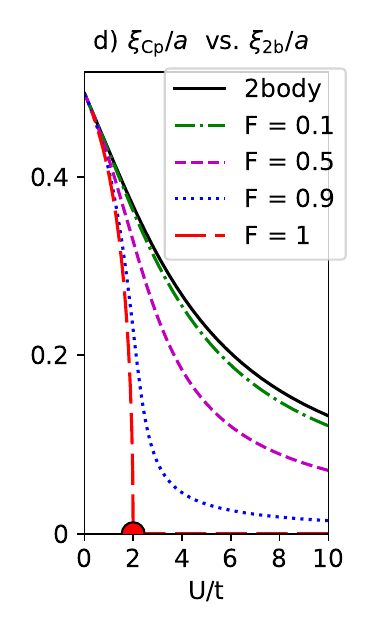}

\caption{\label{fig:xiCp}
The average Cooper-pair size $\xi_\mathrm{Cp}$ is shown in (a) for the Creutz
ladder and in (c) for the $\chi$ lattice as functions of the particle filling
$F$ and the interaction strength $U$. In panels (b) and (d), we compare
$\xi_\mathrm{Cp}$ with the size of the lowest-bound two-body state
$\xi_{2b}$ for the Creutz ladder and the $\chi$ lattice, respectively.
Note that $\xi_\mathrm{Cp}$ reduces to $\xi_{2b}$ in the dilute limit for all $U$.
}
\end{figure*}

In Fig.~\ref{fig:xiCp}, we present the self-consistent solution for
the average Cooper-pair size obtained from Eq.~(\ref{eqn:xiCp}) for the
many-body problem, together with Eqs.~(\ref{eqn:gap}) and~(\ref{eqn:mu}),
as well as the corresponding result from Eq.~(\ref{eqn:xi2b}) for the
two-body problem. We first note that the tensors are diagonal, i.e.,
$
(\xi^2_{2b})_{ij} = \xi^2_{2b} \delta_{ij}
$
and
$
(\xi^2_\mathrm{Cp})_{ij} = \xi^2_\mathrm{Cp} \delta_{ij},
$
and isotropic as a direct consequence of the uniform-pairing condition.

In the dilute limit $F \to 0$, our numerical solutions are fully consistent
with the physically intuitive expectation that $\xi_\mathrm{Cp} \to \xi_{2b}$ for
all values of $U/t$~\cite{iskin25}. 
This coincidence can also be obtained analytically
from Eqs.~(\ref{eqn:xi2b}) and~(\ref{eqn:xiCp}) by noting that
$E_n \approx \xi_n$ and $\mu \to E_b/2$ in the $F \to 0$ limit.
Similarly, in the $U/t \to 0$ limit, we observe that $\xi_\mathrm{Cp} \to \xi_{2b}$
for all values of $F$. This behavior can again be understood 
analytically by noting that
$E_b \to -2\epsilon$ and $E_n \approx \xi_n \ll t$, which leads to
$
\{\xi^2_{2b}, \xi^2_\mathrm{Cp}\} \to \frac{1}{N_c} \sum_\mathbf{k} g_\mathbf{k}
$
in the $U/t \to 0$ limit.
Thus, we expect
$
\{\xi^2_\mathrm{Cp}, \xi^2_{2b}\} \to a^2 / 2
$
for the Creutz ladder and
$
\{\xi^2_\mathrm{Cp}, \xi^2_{2b}\} \to a^2 \chi^2 / 4
$
for the $\chi$ lattice, both of which are in excellent agreement with the
numerical results.

More interestingly, our numerical results show that $\xi_\mathrm{Cp}$ vanishes at
half filling for all $U > U_c$, suggesting an apparent localization of Cooper
pairs to a single lattice site even in the $\Delta_0/t \to 0$ limit. Whether
this behavior is physical or an artifact of the present approximations remains
unclear. In particular, it may indicate that the definition of pair size based
on the localization tensor becomes ill-defined or fails to faithfully
characterize the internal structure of pairs in this regime.
Away from half filling, however, Cooper pairs generally exhibit a finite and
small size. This behavior contrasts sharply with that of conventional BCS
superconductors, where the pair size scales inversely with $\Delta_0$ and
therefore diverges in the $U/t \to 0$ limit. The absence of such a divergence
here highlights the central role of quantum geometry in controlling the spatial
extent of pairing in flat-band systems.

When $\xi_\mathrm{Cp} \to 0$, the Cooper pairs become strongly 
localized in real space, corresponding to purely onsite pairing. 
However, the vanishing of the pair size does not imply the absence of 
superfluidity, since $\xi_\mathrm{Cp}$ characterizes the internal 
structure of pairs rather than their ability to establish phase coherence. 
This is analogous to a superfluid BEC, where the spatial extent of the 
constituent particles, e.g., atoms, does not determine the existence of 
superfluidity. Instead, phase coherence arises from phase rigidity 
established through intersite processes.
While the coherence length $\xi_0$ characterizes amplitude (Higgs-mode) 
fluctuations of the order parameter, superfluid transport is governed by 
the phase stiffness, which depends on the phase dynamics of the condensate. 
Therefore, neither $\xi_\mathrm{Cp}$ nor $\xi_0$ alone provides a direct 
measure of superfluidity. In flat-band systems, a finite superfluid 
weight can arise from quantum-geometric contributions~\cite{peotta15}, 
reflecting the fact that pair mobility is encoded in the structure of 
the Bloch states rather than in band dispersion. Consequently, even 
highly localized pairs can support phase coherence provided that the 
geometry-induced pair mobility, and hence the superfluid stiffness,
remains finite. 
In the regime when $\xi_\mathrm{Cp}$ is finite but the pairing gap 
$\Delta_0$ is large, the system corresponds to tightly-bound pairs, 
analogous to the BEC regime of the BCS-BEC crossover~\cite{strinati18,pistolesi96}. 
Although a large $\Delta_0$ stabilizes pair formation, it does not 
guarantee strong phase coherence. The superfluid response is instead 
controlled by the phase stiffness, determined by the effective mass 
and mobility of the pairs. In flat-band superconductors, this mobility 
is governed by quantum geometry rather than kinetic energy, implying 
that pairing strength and phase coherence arise from distinct mechanisms. 
As a result, even with a large gap, the superfluid response may be 
limited if the geometry-induced pair mobility is suppressed.

\begin{figure*}[htb]
\includegraphics[width=0.25\textwidth,clip=false]{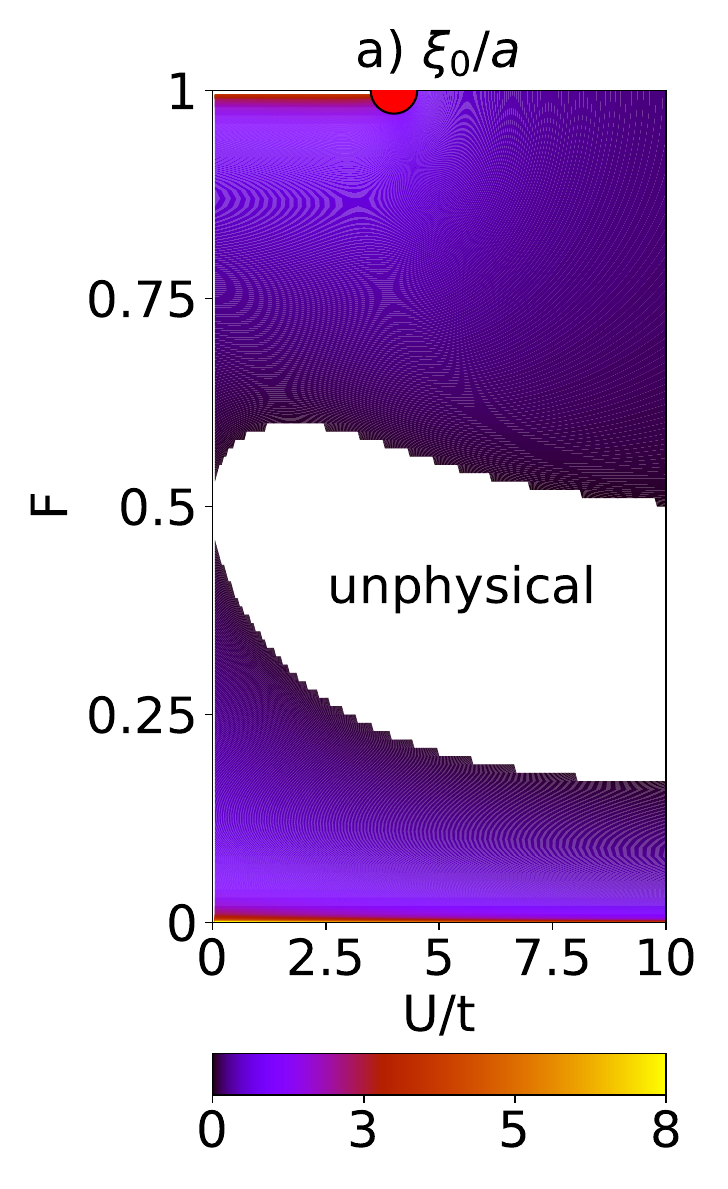}\hfill
\includegraphics[width=0.25\textwidth,clip=false]{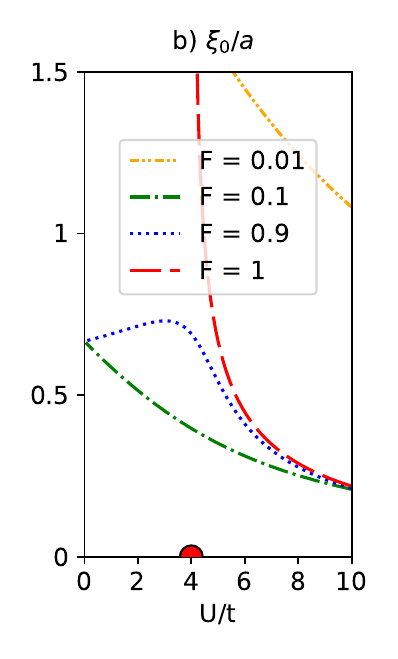}\hfill
\includegraphics[width=0.25\textwidth,clip=false]{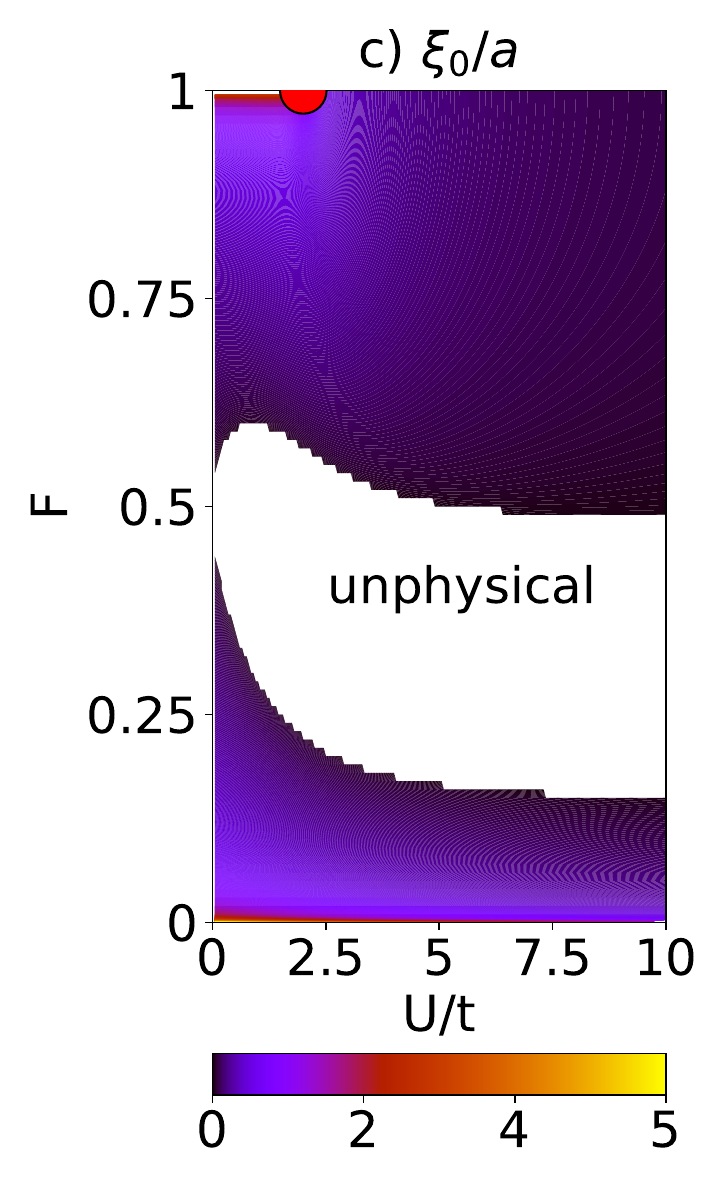}\hfill
\includegraphics[width=0.25\textwidth,clip=false]{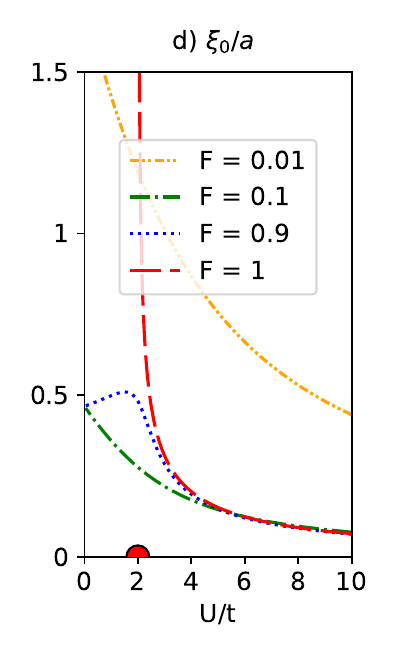}

\caption{\label{fig:xi0}
The zero-temperature coherence length $\xi_0$ is shown in (a) for the
Creutz ladder and in (c) for the $\chi$ lattice as functions of the
particle filling $F$ and the interaction strength $U$. Regions where
$\xi_0^2 < 0$ are shown in white, indicating that the coherence length
is ill defined. In panels (b) and (d), we show representative line cuts
for the Creutz ladder and the $\chi$ lattice, respectively. The coherence
length $\xi_0$ diverges both in the dilute limit and in the vicinity of
the insulating regime at half filling.
}
\end{figure*}

In Fig.~\ref{fig:xi0}, we present the self-consistent solution of
the zero-temperature coherence length obtained from Eq.~(\ref{eqn:xi0}),
together with Eqs.~(\ref{eqn:gap}) and~(\ref{eqn:mu}). Similar to the
pair-size tensors, this tensor is also diagonal and isotropic,
$
(\xi^2_0)_{ij} = \xi^2_0 \delta_{ij},
$
as a direct consequence of the uniform-pairing condition.
The white regions in Figs.~\ref{fig:xi0}(a) and~\ref{fig:xi0}(c)
correspond to parameter regimes in which $\xi^2_0$ becomes negative,
rendering the coherence length ill defined. This issue can be cured
by treating density fluctuations on the same footing as pairing
fluctuations and subsequently performing the low-$\mathbf{q}$
expansion around the new minimum, where the amplitude-amplitude
fluctuation sector occurs at a finite $\mathbf{q} \ne \mathbf{0}$,
rather than at $\mathbf{q} = \mathbf{0}$~\cite{benfatto02}.
Apart from this known subtlety, in the $U/t \to 0$ limit our 
numerical calculations are fully consistent with the analytical 
expectation that
$
(\xi^2_0)_{ij} \to \big[ \frac{1}{F(1-F)} - 4 \big]
\frac{1}{8N_c} \sum_{\mathbf{k}} g^{\mathbf{k}}_{ij}.
$
In the weak-coupling and dilute regime, the leading 
asymptotic behavior is 
$
\xi^2_0 \to \frac{1}{8 F} \frac{1}{N_c} \sum_\mathbf{k} g^\mathbf{k}_{ij}
$, 
showing that the leading divergence is independent of the order in 
which the limits 
$
U/t \to 0
$
and
$
F \to 0.
$
For instance, in the dilute limit $F \to 0$, the coherence length
diverges as
$
\xi^2_0 \to a^2/(16F)
$
for the Creutz ladder and as
$
\xi^2_0 \to a^2 \chi^2/(32F)
$
for the $\chi$ lattice.

To gain further insight into the physical origin of this divergence,
we recast $\xi_0$ in terms of effective pair parameters. To make this connection explicit, we identity an effective mass 
tensor $(M^{-1}_\mathrm{Cp})_{ij}$ for Cooper-pairs, defined from the 
small $\mathbf{q}$-expansion of the pair dispersion, analogous to the 
two-body result in Eq. (\ref{eqn:Mij}). In the dilute limit this 
tensor can be approximated with the lowest two-body bound state,  
$
(M^{-1}_\mathrm{Cp})_{ij} \approx (M^{-1}_{2b})_{ij}
$ 
~\cite{iskin24, iskin24c}.
Using this identification, together with the expressions for the 
coefficients $A$ and $C_{ij}$ in Eqs. (\ref{eqn:A}) and (\ref{eqn:Cij}), 
the coherence length in Eq. (\ref{eqn:xi0}) can be rewritten in terms of 
effective bosonic parameters. In the dilute limit when $U/t \to 0$,
substituting $E_b = -2\epsilon - U/2$ in Eq.~(\ref{eqn:Mij}) gives  
$
(M^{-1}_{2b})_{ij} = \frac{U}{2 N_c} \sum_\mathbf{k} g_{ij}^\mathbf{k},
$
and we obtain
\begin{align}
(\xi^2_0)_{ij} = \frac{(M^{-1}_\mathrm{Cp})_{ij}}{4 U_\mathrm{Cp} F_\mathrm{Cp}},
\label{eqn:xi0bec}
\end{align}
where
$
U_\mathrm{Cp} = 2U
$
is the effective onsite pair-pair repulsion and
$
F_\mathrm{Cp} = F/2
$
denotes the effective pair filling, corresponding to the average
number of condensed Cooper pairs per lattice site~\cite{iskin24, iskin24c}.
Similarly, in the $U/t \to \infty$ limit, we have $E_b \to -U$, which
leads to
$
(M^{-1}_\mathrm{B})_{ij} \approx(M^{-1}_{2b})_{ij} = 
\frac{4 \epsilon^2}{U N_c}
\sum_\mathbf{k} g^\mathbf{k}_{ij},
$ 
where 
$
(M^{-1}_\mathrm{B})_{ij}
$ 
is the effective mass tensor of the composite bosonic pairs,
together with
$
A = F(2-F)/U \to 2F/U
$
and
$
C_{ij} = \frac{2 \epsilon^2}{U^3 N_c}
\sum_\mathbf{k} g^\mathbf{k}_{ij}.
$
This correspondence again reproduces essentially the correct bosonic 
form of the coherence length,
$
(\xi^2_\mathrm{B})_{ij} = (M^{-1}_\mathrm{B})_{ij} / (2 U_\mathrm{BB} F_\mathrm{B}),
$
as expected for a weakly interacting dilute Bose gas, up to a numerical
factor of order unity~\cite{pistolesi96, benfatto02, iskin24c}.  Here $U_\mathrm{BB}$ and $F_\mathrm{B}$ are the effective 
interaction strength and density of the composite bosonic pairs, respectively. 
Thus, although $\xi_0$ becomes unphysical near half filling, 
it correctly reproduces the coherence length in the dilute limit
~\footnote{Away from the dilute limit, in the $U/t \to 0$ limit,
the average number of condensed Cooper pairs per lattice site can be
written as $F(1-F)/2$. One may therefore expect the relevant physical
length scale to take the form
$
(\xi_0^2)_{ij} \to \frac{1}{8F(1-F)N_c} \sum_{\mathbf{k}} g^{\mathbf{k}}_{ij},
$
in accordance with the bosonic expression for the coherence length.}.

Furthermore, our numerical results demonstrate that $\xi_0$ and $\xi_\mathrm{Cp}$ 
are generally of the same order of magnitude, except in the dilute limit 
and near half filling as $U \to U_c$. Specifically, $\xi_0$ diverges in the 
dilute limit for any $U$, as well as at half filling ($F = 1$) as $U$ 
approaches $U_c$. This divergence highlights that $\xi_0$ and $\xi_\mathrm{Cp}$ 
represent two distinct physical length scales, despite their close 
correspondence over a broad range of parameters. Although these quantities 
are physically distinct, they scale identically within weak-coupling 
BCS theory, where both are governed by the ratio of the Fermi velocity 
to the superconducting gap $\Delta_0$. Consequently, both length scales 
diverge in the $U/t \to 0$ limit as $\Delta_0 \to 0$.

Our result on the absence of a BCS-like divergence of characteristic
length scales in flat-band superconductors is consistent with the
recent literature, where the pair size is characterized through the
spatial decay of normal and anomalous correlation functions in real
space~\cite{thumin24}. 
In particular, for the Creutz ladder at quarter filling, the
anomalous correlation function is shown to be strictly finite ranged,
with no characteristic length scale exceeding the lattice spacing.
For the $\chi$ lattice, the anomalous correlation function is found to
be strictly local, implying a vanishing Cooper-pair size. At first
sight, these results appear to contradict our findings. However, this
apparent discrepancy originates from the use of different, though
closely related, definitions of the pair size.

To make this connection explicit, let's consider the anomalous 
correlation function
$
K_{SS'}(\bar{\mathbf{r}}) = \langle c_{S \mathrm{i} \uparrow}
c_{S' \mathrm{i}' \downarrow} \rangle
$
~\cite{thumin24, scirep}.
By transforming this expression first to reciprocal space and then to
the band basis, and by comparing the resulting expression with
Eq.~(\ref{eqn:phi}), one finds
$
K_{SS'}(\bar{\mathbf{r}}) = - A_\mathrm{Cp} \Phi^*(r_{\mathrm{i} S},r_{\mathrm{i}'S'}).
$
Consequently, the Cooper-pair localization tensor can be reexpressed 
as
\begin{align}
(\xi_\mathrm{Cp}^2)_{ij} = \frac{\sum_{\mathrm{i} S \mathrm{i}' S'}
\bar{r}_i \bar{r}_j |K_{SS'}(\bar{\mathbf{r}})|^2}
{\sum_{\mathrm{i} S \mathrm{i}' S'} |K_{SS'}(\bar{\mathbf{r}})|^2}.
\label{eqn:xiK}
\end{align}
In the $U/t \to 0$ limit at quarter filling, one finds 
$
K_{AA}(\bar{r}) = K_{BB}(\bar{r}) = (2 \delta_{\bar{r},0}
- i\delta_{\bar{r},a} + i\delta_{\bar{r},-a})/8
$
and
$
K_{AB}(\bar{r}) = K_{BA}(\bar{r}) = (\delta_{\bar{r},a}
+ \delta_{\bar{r},-a})/8
$
for the Creutz ladder~\cite{thumin24}.
Similarly, the anomalous correlators are given by
$
K_{AA}(\bar{\mathbf{r}}) = K_{BB}(\bar{\mathbf{r}})
= \delta_{\bar{\mathbf{r}},\mathbf{0}}/4
$
and
$
K_{AB}(\bar{\mathbf{r}}) = K_{BA}(\bar{\mathbf{r}}) =
\frac{1}{4 N_c} \sum_{\mathbf{k}} e^{i \mathbf{k} \cdot \bar{\mathbf{r}}}
e^{-i \gamma_\mathbf{k}}
$
for the $\chi$ lattice~\cite{thumin24}. 
Substituting these expressions into Eq.~(\ref{eqn:xiK}) yields
$
\xi_\mathrm{Cp} = a / \sqrt{2}
$
for the Creutz ladder and
$
\xi_\mathrm{Cp} = a \chi/2
$
for the $\chi$ lattice. These results are in perfect agreement with our
analytical predictions and numerical calculations, thereby
demonstrating the consistency between the correlation-function-based
definition of the pair size obtained from the real-space
Bogoliubov-de Gennes formalism and the geometry-based momentum-space
formulation employed in this work.

Furthermore, according to Ref.~\cite{hu23}, when all bands in the 
Bloch spectrum are perfectly flat, the coherence length is predicted 
to be a constant determined by a weighted average of the quantum 
metric. In sharp contrast, we find that neither the coherence 
length nor the pair size remains constant across parameter space. 
The only exception occurs at half filling, where the pair size 
vanishes, signaling strictly local pairing. In the strong-coupling 
limit $U/t \to \infty$, we further observe that the two-body pair 
size, the average Cooper-pair size, and the zero-temperature 
coherence length all scale inversely with $U$, in agreement with 
standard BCS-BEC crossover physics. This behavior demonstrates that, 
within our formulation, these characteristic length scales are not 
bounded from below, in contrast to the quantum-metric lower 
bound proposed in Ref.~\cite{hu23}.
We attribute the origin of this discrepancy to the projection of 
fermionic operators onto the flat band employed in Ref.~\cite{hu23}, 
which effectively restricts the analysis to the $U/t \to 0$ limit. 
Within this restricted framework, the quantity identified as a 
lower bound in Ref.~\cite{hu23} instead emerges as an upper bound 
once interaction effects beyond the flat-band projection are properly 
taken into account. In the present work, we focus exclusively on 
all-flat two-band lattices in order to provide a transparent analysis 
and directly address the recent controversy raised in 
Refs.~\cite{thumin24, scirep}. Extensions to more general band 
structures, including systems with dispersive bands and with or 
without band touchings, are discussed in 
Refs.~\cite{iskin23, iskin24c, iskin25}.

Finally, we comment on the role of the assumptions 
employed in our analysis. The simplifications arising from perfectly 
flat bands, time-reversal symmetry, and uniform pairing allow us 
to obtain compact analytical expressions and to isolate the geometric 
origin of the characteristic length scales.
First, if perfect band flatness is lifted by introducing a small but 
finite dispersion, the characteristic length scales acquire conventional 
intraband contributions~\cite{iskin23, iskin24c, iskin25}. 
In the weak-coupling limit, the usual BCS coherence length, set by 
the Fermi velocity, will eventually dominate and diverge. Nevertheless, 
the interband geometric contributions derived here are expected to 
persist, providing a finite contribution to the spatial extent of 
Cooper pairs. Second, relaxing the uniform-pairing condition, e.g., 
through sublattice-asymmetric potentials or disorder, renders the 
mean-field order parameter spatially dependent. In this case, 
the direct relation between the characteristic length scales and the 
quantum geometry of the Bloch states becomes more involved and 
requires further investigation.
Finally, breaking time-reversal symmetry can qualitatively modify the 
geometric contributions. For example, in topological flat bands, 
the pairing geometry can become intertwined with Berry curvature 
effects, leading to additional contributions beyond the quantum metric. 
In all these scenarios, by analogy with the superfluid BEC of atoms, 
we expect the general expression in Eq.~(\ref{eqn:xi0bec}) to remain 
valid within the flat-band regime, albeit with renormalized and more 
intricate effective parameters for the Cooper pairs. Elucidating this 
interplay remains a compelling direction for future research, 
toward which we have already made some progress~\cite{keskiner26}.

\section{Conclusion}
\label{sec:conc}

In summary, we systematically examined characteristic length scales
associated with pairing and coherence in superconducting systems with
perfectly flat bands from three complementary perspectives. First, we
analyzed the localization tensor of the lowest-lying two-body bound
state. Second, we studied the average size of Cooper pairs within the
mean-field approximation. Third, we investigated the zero-temperature
coherence length within the Gaussian-fluctuation theory. The presence of
time-reversal and sublattice-exchange symmetries, which enforce 
spatially-uniform pairing, allowed us to substantially simplify 
the analysis and to cleanly isolate effects arising purely from 
quantum geometry.

Our results demonstrate that, throughout the parameter space, both the
two-body pair size and the many-body Cooper-pair size, unlike in
conventional superconductors with dispersive bands, do not exhibit a
BCS-like divergence in the weak-coupling limit. Instead, these length
scales remain finite and small, and are entirely governed by the quantum
geometry of the underlying Bloch states. In particular, in the
weak-coupling regime both pair sizes reduce to Brillouin-zone averages
of the quantum metric, highlighting the central role of band geometry
when kinetic energy is quenched.
The zero-temperature coherence length displays a related geometric 
origin, except in the dilute regime and in parameter regions 
that are insulating or proximate to an insulating phase, where 
its behavior becomes qualitatively distinct. This contrast 
highlights that the coherence length and the pair size encode 
fundamentally different physical information: while the pair size 
reflects the internal spatial structure of bound fermion pairs, 
the coherence length is governed by collective properties and 
critical fluctuations of the superconducting state~\cite{iskin24}. 
Taken together, our findings clarify the geometric origin of 
pairing length scales in flat-band superconductors and underscore 
the necessity of distinguishing between pair size and coherence 
length in such systems.

Moreover, our results demonstrate that the apparent 
discrepancies in the recent literature originate from comparing 
different definitions of superconducting length scales that probe 
distinct physical properties. Our findings are consistent with 
Ref.~\cite{thumin24}, where correlation-function-based approaches 
show that Cooper pairs remain short-ranged in flat-band systems. 
By introducing the average Cooper-pair size within a momentum-space 
(localization-tensor) framework, we make explicit that this length 
scale is governed by the quantum metric, thereby providing a direct 
geometric interpretation of these real-space results within a 
unified formalism.
By contrast, our results differ from Ref.~\cite{hu23}, where the 
coherence length was interpreted as a geometry-controlled and 
finite quantity. Within our framework, which treats pairing and 
collective fluctuations on equal microscopic footing, we find that 
the coherence length exhibits a qualitatively different dependence 
on interaction strength and filling, including divergences in the 
dilute limit and near insulating regimes. This demonstrates 
that the behavior identified in Ref.~\cite{hu23} does not capture 
the full parameter dependence of superconducting correlations.
At the same time, our analysis clarifies the origin of the proposed 
lower bound on the coherence length in Ref.~\cite{hu23}: 
it arises from the projection onto a flat band and is therefore 
restricted to the weak-coupling regime. When interaction effects 
beyond this approximation are properly included, this bound does 
not persist and instead corresponds to an upper bound in the 
weak-coupling limit. Taken together, our results provide a 
unified framework that reconciles these earlier approaches by 
showing that pair size and coherence length are distinct quantities 
governed by quantum geometry in fundamentally different ways.

Looking ahead, it would be valuable to explore how the geometric 
control of pairing length scales identified here evolves in more 
general settings, including weakly dispersive bands, multiband 
systems without sublattice-exchange symmetry, and treatments that 
incorporate beyond-mean-field corrections. 
In particular, it would be interesting to investigate to what extent 
these geometric effects persist in strongly-correlated regimes using 
numerically exact methods such as DMRG, where the relevant length
scales could be extracted from appropriate many-body correlation 
functions~\cite{mondaini18, chan22}.

\begin{acknowledgments}
We acknowledge support from the U.S. Air Force Office of Scientific
Research (AFOSR) under Grant No.~FA8655-24-1-7391.
\end{acknowledgments}

\bibliography{refs}

\end{document}